# Systematically Evaluating Equivalent Purpose for Digital Maps

### Table 1: Authors

| Name | Institution | Email |
|---|---|---|
| Brandon Biggs | Georgia Institute of Technology and Smith-Kettlewell Eye Research Institute | brandon.biggs@ski.org |
| David Sloan | Vispero | dsloan@tpgi.com |
| Brett Oppegaard | University of Hawaiʻi at Mānoa | brett.oppegaard@hawaii.edu |
| Nicholas A. Giudice | University of Maine | nicholas.giudice@maine.edu |
| James M. Coughlan | Smith-Kettlewell Eye Research Institute | coughlan@ski.org |
| Bruce N. Walker | Georgia Institute of Technology | bruce.walker@psych.gatech.edu |

### Abstract

Digital geographic maps remain largely inaccessible to blind and low-vision individuals (BLVIs), despite global legislation adopting the Web Content Accessibility Guidelines (WCAG). A critical gap exists in defining "equivalent purpose" for maps under WCAG Success Criterion 1.1.1, which requires that non-text content provide a text alternative that serves the "equivalent purpose". This paper proposes a systematic framework for evaluating map accessibility, called the Map Equivalent-Purpose Framework (MEP Framework), defining purpose through three items (Generalized, Spatial Information, and Spatial Relationships), and establishing 15 measurable criteria for equivalent information communication. Eight text map representations were evaluated against visual map baselines using the proposed MEP Framework. Results show that legacy methods such as tables and turn-by-turn directions fail to meet the MEP Framework criteria, while Audiom Maps, Multi user Domain (MUD) Maps, and Audio Descriptions meet the criteria. The evaluation highlights the necessity of holistic, systematic approaches to ensure non-visual maps convey all generalized spatial information and relationships present in visual maps. The MEP Framework provides a replicable methodology for comprehensively assessing digital map accessibility, clarifying WCAG's equivalent purpose, and guiding compliant and usable map creation. Compliant maps will support BLVIs' participation in map-dependent professions and civic engagement.

## Introduction

While the *Web Content Accessibility Guidelines (WCAG) Overview* are becoming part of global digital accessibility legislation (U.S. Department of Justice Civil Rights Division; W3C Web Accessibility Initiative ), ambiguity exists, particularly around digital geographic map accessibility (Biggs, Coughlan, et al.; Sloan; KMSOC). This paper proposes a systematic definition and evaluation framework, called the Map Equivalent-Purpose Framework (MEP Framework), in order to determine if blind and low vision individuals (BLVIs) can equally obtain the same information as their sighted counterparts from digital geographic maps. WCAG success criterion (SC) 1.1.1 states: "non-text content that is presented to the user has a text alternative that serves the equivalent purpose", but what does "*equivalent purpose*" mean for geographic maps (World Wide Web Consortium)? According to the World Wide Web consortium (W3C) technique G92, there is a three-part test to determine equivalent purpose for WCAG SC1.1.1: "(1) Remove, hide, or mask the non-text content; (2) Display the long description; and (3) Check that the long description conveys the same information conveyed by the non-text content" (Accessibility Guidelines Working Group). Note although the longDesc element is deprecated in HTML, the following tutorial outlines other methods of providing long descriptions, including by using the figcaption element and or a link to another page (W3C Web Accessibility Initiative). The intent behind SC1.1.1. is clear, but with the numerous existing interpretations of this criterion around maps, a structured evaluation framework is needed (Minnesota IT Services; Sloan; Logan, *Accessible Maps on the Web*). The proposed MEP Framework consists of a three-part framework for defining purpose, and a 15-part framework for defining equivalent information communication. For an equivalent purpose to be achieved-as compared to a non-text map baseline-all the criteria from both parts of the MEP Framework need to be satisfied.

*hTable 2: Text Map Techniques*

*Table2 shows Common text map techniques from the web accessibility community, which emphasize qualitative experience or primary purpose rather than WCAG-defined equivalent purpose, alongside empirically tested and co-designed research-based alternatives*

| Technique | Category | Typical Use / Description | Key References |
|---|---|---|---|
| Turn-by-turn directions | Web Accessibility Community | Sequential navigation instructions describing how to travel between locations | (Sloan) |
| Nearby address search | Web Accessibility Community | Returns addresses or points of interest near a specified location | (*PO Locator*; *AirNow*) |
| List of locations | Web Accessibility Community | Linear listing of places shown on a map | (Shull; Logan, *Accessible Maps on the Web*; Juan-Armero and Luján-Mora) |

| Technique | Category | Typical Use / Description | Key References |
| --- | --- | --- | --- |
| Alternate text | Web Accessibility Community | Short textual summary describing the map or its primary message | (Minnesota IT Services; McCall and Chagnon; Cassidy) |
| Tables | Web Accessibility Community | Tabular presentation of map data values | (Sloan; *United States COVID-19 Cases, Deaths, and Laboratory Testing (NAATs) by State, Territory, and Jurisdiction*) |
| Interactive alternate text | Research-based map technique | Spatially navigable text interface where users explore nearby or intersecting features | (Biggs, Toth, et al.; Hennig et al.; Loeliger and Stockman; Zhao et al.; Biggs, Yusim, et al.) |
| Audio description | Research-based map technique | Long-form, structured textual description often | (Conway et al.; UniDescription) |

| Technique | Category | Typical Use / Description | Key References |
| --- | --- | --- | --- |
| | | spanning multiple pages | |
| Chat interfaces | Research-based map technique | Conversational querying of spatial information via natural language | (Froehlich et al.; Jain et al.) |

Despite these methods, most major map tools fail to provide any text alternative by default (Biggs, Coughlan, et al.). The text representation needs to allow the user to generate their own routes, perform spatial modeling to understand "What is nearby and why," make decisions based on multiple layers of information, understand what spatial relationships are not present in the data, and comprehend spatial information at a professional level (Arundel and Li). There are non-visual sensory map representations, such as raised-line tactile maps (Biggs, Pitcher-Cooper, et al.), 3D model maps (Coughlan et al.), vibro auditory maps (Giudice et al.), and auditory maps (Loeliger and Stockman), but WCAG requires a digital keyboard accessible text map.

There are hundreds of definitions of a "map," most of which are vision-centric, but at a map's most fundamental level, all these definitions agree that the purpose of a map is to communicate "generalized spatial information and relationships" (Lapaine et al.; *Map Definition*; *Map Definition*). "Generalized" means symbols are being used to approximately represent the real-world (e.g., as Belgian artist René Magritte asserted in "The Treachery of Images," "Ceci n’est pas une pipe” (“This is not a pipe”) (Harvey)). Common symbols on maps include points, lines, and polygons (White). “Spatial information” is about where features are, and what they are like (e.g., shape, size, orientation, location, and other properties) (Li and Huang; Arundel and Li). “Spatial relationships” mean that location, distance, topological intersections and adjacencies, and direction between objects are core to understanding the information in the representation (Arundel and Li; Clementini et al.). Representation in this context is a digital depiction of data. This paper and definition focuses on two-dimensional geographic maps, but other spatial representations (e.g., mind maps, anatomical diagrams, and three-dimensional diagrams) may also fit this definition. This map definition is meant to be fundamental to every two-dimensional geographic map, no matter the subpurposes it may have (e.g., historical, navigational, or analytical). Since the fundamental purpose of a map is to communicate generalized spatial information and relationships, equivalency could be determined by assessing the quality and comprehensiveness of these three areas.

Information communicated via maps can be categorized under such labels as landmark, route, and survey knowledge (Siegel and White). Using the implicit variables provided by Siegel and White, and other cartographic elements, this paper proposes a set of measurable variables and an assessment that can be used to evaluate the communication of landmark, route, and survey knowledge (Biggs, Coughlan, et al.). Landmark knowledge includes the feature-specific cartographic variables of shape, size, orientation, and temporal information (Roth; White; Song), as well as feature name and type (Guidero), and the numeric and or categorical variables used in thematic maps (Foster). Survey knowledge includes distance, direction, topology, and location relationships of all features on a map (e.g. global configuration (Arundel and Li; Li and Huang; Clementini et al.)). Route information is not present on all maps, but it consists of how to get between two or more features on a map following a particular defined path, including the landmark and survey elements for the designated route object, and that the route information is as prevalent in the text map as it is on the visual map (Arundel and Li; Aziz et al.). With such an approach, these fundamental and measurable variables should be able to convey all generalized spatial information and relationships present in a map representation to determine equivalency (e.g., density using distance, size, topology, color, shading, and shape). The evaluation in this paper will use the equivalent purpose variables provided in these last two paragraphs to determine if different representations meet WCAG SC1.1.1 standards.

**Table 3: Map Equivalent-Purpose Framework Overview**

The final MEP Framework template, the guide to use the template, and the final evaluation with majority agreed-on variables can be found in the supplemental data.

| Type | Spatial Knowledge Variable | Row Description | Example of Top Rows, and Passing Evaluation Area |
| --- | --- | --- | --- |
| Map Representation Description | | Representation Description is a description of the representation being evaluated using this framework. The descriptions will all be in this row. This is not a description of the contents of the map, but a description of how the representation map being evaluated works. | A table has rows for all objects on the map, and a column for the numeric or categorical object properties. |
| Text Map | | This is the text version of the map that is being evaluated against the visual map. It is important to keep in mind that the text map is being compared with the visual map, and not being evaluated on its own. | https://purvasingh96.github.io/accessible-maps/#/tables/table1 |

| Visual Baseline Map | | The Visual Map Comparison is a visual only map that the representation in the row above is being evaluated against. This is the baseline representation. It is the map as originally represented, via original file, screenshot, or image, if analog. | https://purvasingh96.github.io/accessible-maps/#/maps/map1 |
|---|---|---|---|
| Total Purpose Items Passed: | 0/3 Items Passed | The purpose of a map is to communicate generalized spatial information and relationships. These are the three components of a map at its most fundamental level. If any of these three elements are missing, as compared with a visual comparison, then the purpose of the map is not being met. You should add a fraction in this row, the top number is the passing elements, and the bottom number is the total elements. This is the first quick evaluation to identify if further evaluation is needed.<br>Note: if the area is not present on the visual baseline map, it does not need to be evaluated on the text map. | 2/3 Items Passed |
| | Generalized: | General means that "something" is being used to represent the real-world. The primitive map symbols are points, lines, and polygons. Are all the points, lines, and polygons for every object on the visual map, described in detail on the text map? All objects on the visual map should be present on the text map. | Johnson Hall is a rectangle two feet long and five feet high. It has a point indicating a door on its southwest corner, facing west. |
| | Spatial Information: | Spatial means that shape, size, and orientation are described in detail for each object present on the text map. Is the shape, size, and orientation described for each object on the text map? | Johnson Hall is 30 feet west (3 o'clock) from the Rec Center. They are both located on the Wonderful University campus, and the Rec Center is at the very northwest corner of the campus. |
|  | Spatial Relationships: | Relationships are the connections between multiple objects. Is there a way to understand the distance, direction, topological, and relative location of every object on the text map between every other object on the text map? Is it possible to understand a line connecting each object on the text map? | Johnson Hall is 30 feet west (3 o'clock) from the Rec Center. They are both located on the Wonderful University campus. |
| Total Equivalent Items Passed: | 0/15 Items Passed | This is a fraction summing all three spatial knowledge areas. the top number is the passing items, and the bottom number is the total number of applicable items. If there are items that are not applicable, they can be left out the total number of items on the bottom. Sum all landmark knowledge, route, knowledge, and survey knowledge areas.<br>Note: if the information is not present on the visual baseline map, it does not need to be evaluated on the text map and should be marked N/A. | 2/15 Items Passed |
| Landmark knowledge: | 0/8 Items Passed | Landmark information is the individual characteristics of every object on the map. These areas should only evaluate the objects present on the text map against the visual baseline map. Go through each object on the text map and evaluate each landmark area against the same object on the visual map. | 8/8 Items Passed |
|  | Sensory characteristics: | These are the colors, sounds, visuals, smells, and other characteristics that are communicated with the user. An example would be a “blue” color being used to represent water on the visual map. This information needs to be described on the text map. The map key is where this information is often provided. | Water is represented through a corn blue color, grass is represented through a dark green color. |

| | Name: | This is the label of the object, or how the object is referenced when discussing the information on the map. For example, a road could be named: “Fillmore St.”. Every object in the text map needs to have a name of some kind. | Fillmore St. |
|---|---|---|---|
| | Type: | The type is a categorical descriptor for the object. For example: “Street”, “Walkway”, “Restaurant”, etc. This could also be derived from the name (e.g., “Fillmore St.” the “St.” means “Street”, which is the type). | Street |
| | Shape: | The shape consists of the detailed borders of an object. The borders of each object on the visual map need to match the borders or shapes described on the text map. | A simple description could be: “Rectangle”, but an irregular complex shape requires more description, for example: “Idaho’s northern border is relatively short, about 40 miles, confined to the narrow panhandle, and is defined by a straight line that separates it from British Columbia, Canada. The eastern border with Montana and Wyoming starts in the north with Montana. For around 100 miles with a 5:30 o'clock slope, the border is fairly strait going north to south with a very gentle slope east. A stronger eastern curve going from 5:30-3 o'clock into Montana happens, and Idaho eventually has Montana completely to the north for about 150 miles until it reaches Wyoming. The eastern border between Idaho and Wyoming is strait for around 100 miles. The southern border is strait for around 479 miles (its widest point). The southern border is split equally between Utah in the east and Nevada in the west. The eastern border between Oregon and Washington is strait for around 305 miles, and 60% in the south is the border with Oregon and the 40% in the north is with Washington.” |
| | Orientation: | The orientation can be combined with the shape description, but should consist of the border objects and how the object relates to those objects, and how the shape is facing. The orientation of every object on the visual map needs to be described on the text map. | "The Blue Tavern is a triangular building with the point of the triangle facing northwest at 11 o'clock. The point touches Fillmore St. To the northeast side of the building is Freedom Parks, to the south is Green Laundromat, and southeast, along the flat side of the triangle, is a patio for the bar." |
| | Size: | The size are the particular dimensions of the perimeter of the object. The size that is shown for every object on the visual map needs to be described on the text map. | "Blue tavern has 3 walls. The two long walls that connect at a point are 50 meters long, and the short wall leading on to the patio is 20 meters long." |
| | Temporal: | Temporal information is the change in data (geographic, numerical, or categorical) over time. If it is possible to observe data change over time on the visual map, then the same time change information needs to be described on the text map. If it is not possible to view different data from different times on the visual map, then write "N/A" in this section. You should be looking for timelines, line graphs, or settings to change the date. | The number of total COVID cases over the U.S. from 1/1/2021 to 4/1/2021:<br>1/1/2021: 500020<br>2/1/2021: 502000<br>3/1/2021: 512000<br>4/1/2021: 523110 |
| | Overlaid information for thematic maps: | Overlaid information for thematic maps consist of the one or more numeric or categorical variables associated with each object. These are normally additional properties attached to each object on the map. This will be different categorical variables than "type", since type is its own item. The overlaid information is normally statistics (e.g., population data). If it is possible to observe different overlaid information on the visual map, then that same data needs to be described in the text map. | California: 543 Total Cases, 5 Total Deaths, and Yes masks Required. |

| Route: | 0/3 Items Passed | Route information are the defined routes on the map. These are normally navigation routes from point A to point B following some kind of path. If there are defined routes present on the visual baseline map that are not present on the text map, all areas in this section fail. Compare the routes on the text map with the routes on the visual baseline map. If there are no defined routes, this entire section should be N/A. | N/A |
|---|---|---|---|
| | Landmark Information: | Make sure all applicable areas of the landmark knowledge section pass with the object representing the defined route. Take into consideration distance markers, shape, length/size, orientation, color, etc. If any items fail from the landmark section for any of the routes, write the entire section as a "No". | There is a blue line tracing the shape of Fillmore St. Every 75 feet there is a blue marker showing the length. There is a blue marker 2/3rds of the way along the line. The line is strait going south to north and represents 94 meters. The line starts at 3321 Fillmore St. and ends at 1283 Fillmore St. The line angles from south to north at a 2 o'clock angle. |
| | Survey Information: | Make sure all applicable areas of survey knowledge are evaluated on each defined route in relationship with the other objects on the map. These other objects include intersections, intersection types, buildings, etc. If any items fail from the survey section for any of the routes, write the entire section as a "No". | The line starts at 3321 Fillmore St. and ends at 1283 Fillmore St. The line begins at the center of the 3321 Fillmore St. building, and continues 10 meters 3 meters in front of the building on the west side of Fillmore St. The route then crosses Clay St. for 4 meters with a stoplight intersection. On the northside of Clay and Fillmore the route continues for 5 meters in front of Dunkin Donuts, 10 meters along Lululemon, 20 meters along Wells Fargo bank, 10 meters in front of a parking lot, 5 meters past Boba and more, 20 40 meters in front of a Safeway, and 10 meters along the front of 1283 Fillmore St., ending half-way along 1283 Fillmore St. |
| | Prominence: | The defined route information should be easy to find and chunked in its own section. The time it takes to find the route information on the visual map should be around the same time it takes to find the | Routes:<br>Route 1: 3321 Fillmore to 1283 Fillmore<br>There is a blue line tracing the shape of Fillmore St. Every 75 feet there is a blue marker showing the length... |
| | | information on the text map, although the text map will probably be a little slower because text takes time to read. | |
| Survey: | 0/5 Items Passed | Survey information is the overall understanding of how every object is related spatially to one another. If there are objects present on the visual baseline map that are not present on the text map, then all items in this section fail. Go through each object on the text map and evaluate if it communicates the same survey information as the same object on the visual baseline map. | Relative Location example:<br>The Red room is a square in the center of the map. The bottom right corner of the square green room is 23 meters and 6 o'clock from the red room. The bottom right corner of the triangular Yellow Room is 10 meters and 2 o'clock from the red room, and the bottom edge of the yellow room is parallel with the top edge of the Red room. The top left corner of the rectangular Orange room is 5 meters and 4 o'clock from the Red room. |
| | Distance between all points, polygons, and lines: | Distance is how far each object is from another, typically using Euclidean distance. This is using the “as the crow flies” distance, not taking into account objects between the starting point and the ending point. If this distance information is present on the visual map, it needs to also be present in the text map. | The green room is 23 meters from the red room. The Yellow Room is 10 meters from the red room. The Orange room is 5 meters from the Red room. |
| | Direction between all points, polygons, and lines: | Direction is the angle between each object. For example: Fillmore St. is 85 degrees from Clay St., or using more imprecise methods if the map is not as complex (e.g., Sandwiches and More is 2 o'clock from Men’s Haircuts; or The Dry Cleaner’s is right of Shoes and More; or Pet Grooming is east of The Crabcake House). The important aspect is that the angle between each object shown on the visual map is accurately communicated in the text map. | The green room is 6 o'clock from the red room. The Yellow Room is 2 o'clock from the red room. The Orange room is 6 o'clock from the Red room. |
| | Absolute location of all points, polygons, and lines: | Are the exact coordinates for the object present?<br>There are two types of location: absolute (with coordinates), and relative (how objects relate to each other). If the map provides access to absolute location, then that should be present at an equal level to the baseline map. Additionally, the relative location between every object needs to be provided through a combination of distance and direction in the other survey areas. If the baseline map shows graticules, the distance and direction between the different graticules and the edges of each shape should be provided. If coordinates are given through a number, then the same coordinates should be provided in both representations. If no absolute location information is present in the baseline map, indicate N/A. If the baseline map and the map being evaluated contain absolute information for the same objects, then write “Yes”, otherwise, write “No”. | Absolute location example:<br>The right edge of the 5 meters h and 10 meters w rectangular Red Room is 10 meters west of the 9 degrees E graticule, and the bottom edge of the Red Room is 150 meters north of the 45 degrees N graticule. The center of the Red Room is at 45.0013722° N, 8.9998094° E. |
| | Topological information of all points, polygons, and lines: | Topological information is the relationship between two objects across three dimensions that can be expressed in a three by three matrix for both objects: interior, boundary, and exterior. If two objects intersect, this intersectionality needs to be detailed across these three dimensions. If one of the dimensions is missing, then indicate “No” in this cell. If there are only bordering objects, then those borders need to be indicated. If objects are all spread apart, then the exterior relationship should be indicated in the distance and direction variables in the other survey information. | The square purple room is placed completely inside 10 meters from the top and four meters from the left edge of the rectangular Colored Room Zone, and the top of the purple room is parallel to the top of the Colored Room Zone.<br>The elevator point is outside, 10 meters above and six meters to the right of the top left corner of the Colored Room Zone.<br>The road of discovery is a line that travels from the elevator, strait down into the Colored Room Zone. It then continues one meter to the right of the purple room, and after 14 meters from the top of the Colored Room Zone, at the edge of the purple room, makes a 90 degree turn to the right… |

*Method*

**Table 4: Evaluated Text-Based Map Representations and Visual Map Sources**

Table 4 Summarizes the eight evaluated text-based map representations, including their sources, visual map baselines, origin type, and supporting literature.

| Text Map Representation | Origin | Text Map Source / Basis | Visual Map Source | Basis / Implementation | Key References |
|---|---|---|---|---|---|
| **Turn-by-Turn Directions** | Existing website | Google Maps directions, commonly suggested as a text map alternative | Same Google Maps link | Directions suggested as a text map alternative; text and visual map co-located | (Sloan; *ClickAndGo Wayfinding*) |
| **Table Map** | Research team (based on existing example) | Tabular map alternative based on CDC COVID-19 heatmap | CDC COVID-19 visual map | Derived from a tabular alternative proposed for thematic maps in prior work | (*United States COVID-19 Cases, Deaths, and Laboratory Testing (NAATs) by State,* |

| Text Map Representation | Origin | Text Map Source / Basis | Visual Map Source | Basis / Implementation | Key References |
| --- | --- | --- | --- | --- | --- |
| | | | | | *Territory, and Jurisdiction*; Sloan; Colorado OIT-GIS) |
| **Nearby Address Search** | Existing website | Bank of America locator “nearby search” interface | Same website map | Common locator pattern used by major map tools and described in literature | (Bank of America Corporation.; *mapbox-gl-accessibility*; Logan, *Accessible Maps on the Web*; Logan, *Let’s Make Maps Widely Accessible*; Juan-Armero and Luján-Mora) |
| **Short Text Alternative** | Research team | Short alt text (<140 characters) following | Google Maps of | Based on guidance limiting alternate text to | (McCall and Chagnon; Desrosiers) |

| Text Map Representation | Origin | Text Map Source / Basis | Visual Map Source | Basis / Implementation | Key References |
| --- | --- | --- | --- | --- | --- |
| | (based on guidance) | common web practices | described location | under 140 characters | |
| **Google Maps Interactive Alternate Text** | Existing website | Google Maps built-in interactive alternative text | Same Google Maps link | Built-in interactive text alternative provided by Google Maps | (Google) |
| **Audiom Map Interactive Alternate Text** | Existing research system | Audiom interactive text map | Same Audiom map | Interactive spatial text map with keyboard-based navigation | (Biggs, Toth, et al.; *Welcome to XR Navigation!*) |
| **MUD Map Interactive Alternate Text** | Research team (based on literature) | Text-adventure style map using rooms and cardinal navigation (Evennia MUD) | Pacific Northwest COVID-19 visual map | Text-adventure-style “rooms” navigated via cardinal directions | (Griatch; The Mud Connector; Williams; Biggs, Yusim, et al.) |

| Text Map Representation | Origin | Text Map Source / Basis | Visual Map Source | Basis / Implementation | Key References |
|---|---|---|---|---|---|
| **Audio Description** | Research team (based on guidelines) | Long-form structured audio description written in a word processor | Pacific Northwest COVID-19 visual map | Long-form structured description authored following audio description guidelines | (*UniDescription Academy*; Conway et al.) |

The MUD Map and Audio Description texts were written with the above MEP Framework in mind. The reason for keeping the framework in mind while writing the text was because it is otherwise easy to leave out details from the map text, but it is also easy to add those details in if they are missing. Many example descriptions were reviewed that follow the *UniDescription Academy* guidelines, but often there were small details missing (e.g., the shape of one object was not described, or size information was missing for a few objects). Since it was the technique being evaluated for this paper rather than an individual description, the research team created their own description, systematically considering all elements of the MEP Framework. The MUD Map and Audio Description text were the same, the difference between the two maps was in the text navigation method. MUD maps split text into "rooms" of varying size that are navigated to by typing cardinal directions (e.g., n, s, e, w) and each "room" has its own text description. Audio descriptions are long-form text indexed by headings and paragraphs rather than rooms.

Using the MEP Framework, eight textual map representations were independently evaluated against a visual map by three members of our research team. After the initial evaluation by one researcher, the definitions and evaluation were refined by the team, and two new researchers, with no prior knowledge of the topic, were brought in to independently evaluate the text representations using the MEP framework. They were provided with the visual map, non-visual map, the above template with definitions, and a guide on how to use the template. Once the two new researchers completed their first round of evaluations, intercoder reliability was 35% using the average Cohen's Kappa. The three researchers then met and discussed the definitions for each area, and refined the definitions of each area and for the text representations for a second evaluation round. After the second round of evaluations the intercoder reliability was 69%. The head researcher then went into each evaluation, and reviewed the areas that the reviewers were differing on and asked for clarification, to resolve contradictions, and for the reviewers to update sections that did not match the provided definitions at all. In the third round, intercoder reliability was 91%.

The evaluation consisted of the researcher reviewing first the non-visual map, followed by the visual map. Rather than evaluating each map directly against the criteria, the two maps were directly compared. This comparison is critical, because if the visual map lacked functionality, then the area would be considered N/A, instead of a fail. If there was no way to view coordinates on the visual map, but the text map did have coordinates, the answer would be "N/A". Although this is not required to effectively use the MEP Framework, a justification was provided for each pass or fail decision. If there was a partial pass, the ultimate score was a fail, but with a note of the partially passing element. Above each section is the sum of passing variables divided by the total number of variables in the section (e.g., 2/3 means 2 out of 3 purpose variables passed).

The three nominal purpose variables defining a map (generalized, spatial, and relationship, described above) were evaluated first. These variables required a conceptual evaluation of the representation. The equivalency variables were evaluated second, and had a total of 15 criteria, although one criterion (overlaid information) was only applicable to thematic maps. These criteria were a mixture of nominal, ratio, continuous, and circular variables.

### *Results*

#### **Figure 1: Bar Chart Showing Scores**

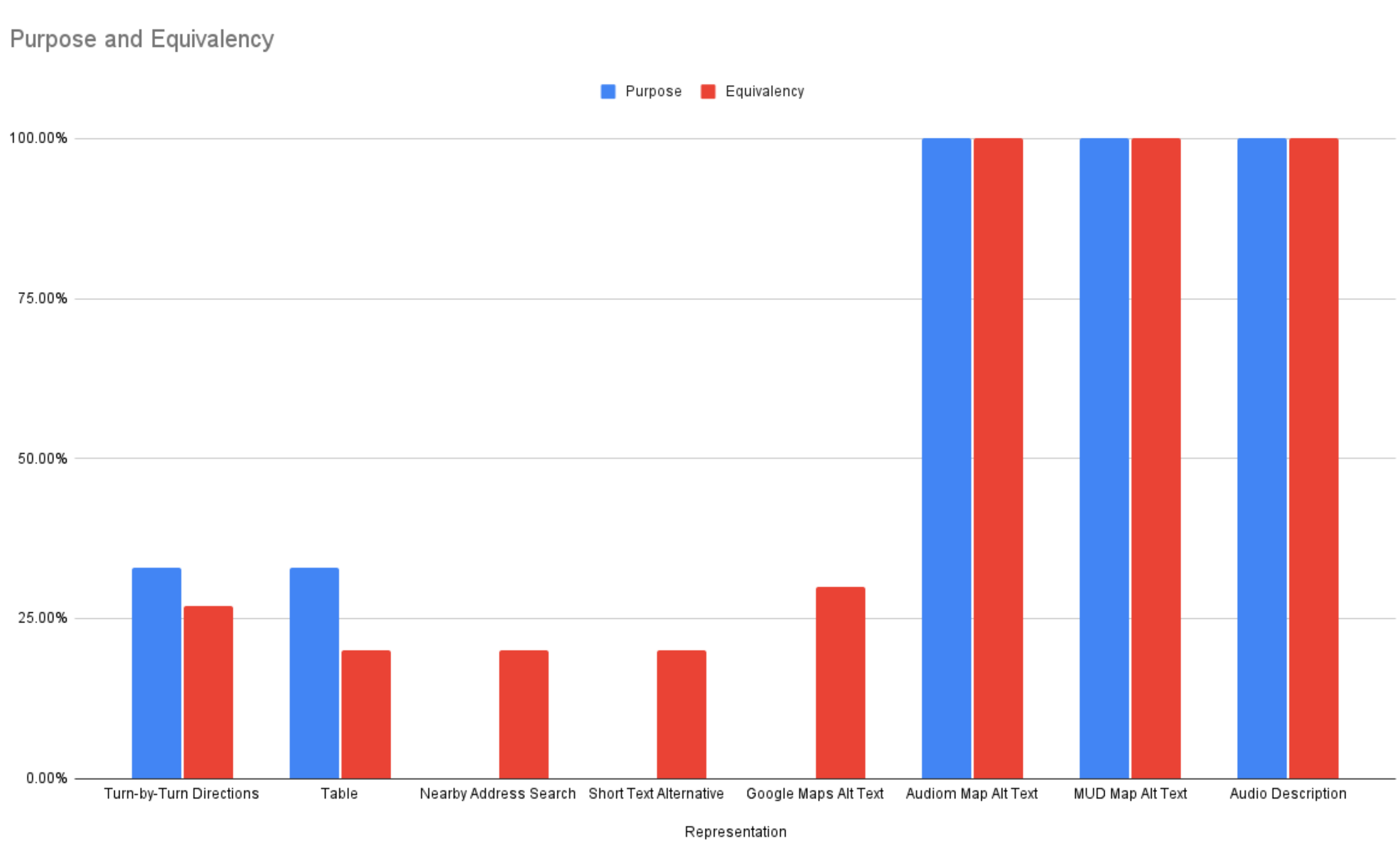

**Table 5: Table Showing Scores**

| **Text Map** | **Purpose** | **Equivalency** | Final Result |
|---|---|---|---|
| tTurn-by-Turn Directions | 33.00% | 23.00% | Fail |
| Table | 33.00% | 20.00% | Fail |
| Nearby Address Search | 0.00% | 20.00% | Fail |
| Short Text Alternative | 0.00% | 20.00% | Fail |
| Google Maps Alt Text | 0.00% | 30.00% | Fail |
| Audiom Map Alt Text | 100.00% | 100.00% | Pass |
| MUD Map Alt Text | 100.00% | 100.00% | Pass |
| Audio Description | 100.00% | 100.00% | Pass |

Out of eight map representations, three (Audiom Map Interactive Alternate Text, MUD Map Interactive Alternate Text, and Audio Description) passed the MEP Framework evaluation and showed they served the equivalent purpose to their corresponding visual map. When representations failed, a maximum of one out of three items passed the purpose test, and three out of fifteen items passed the equivalency test. Moreover, if a representation failed the purpose test, it consequently failed the equivalency test. The difference between passing and failing representations was substantial in the equivalency test: passing maps achieved 100% equivalency, whereas failing maps averaged 23%, producing an approximate 77-point performance gap.

The largest discrepancies were observed in spatial information and survey knowledge-variables that depend on communicating shape, size, orientation, distance, and direction between features. Missing objects in the text maps were the primary cause of failure; four of the five failing maps lacked the same number of features as the visual map. When features were missing from a text map, all areas of survey knowledge automatically failed, since the core requirement for evaluating survey knowledge involved assessing relationships among all points, polygons, and lines on the map.

Landmark knowledge showed the smallest gap across representations (an average of 39% for failing maps versus 100% for passing maps). Most representations successfully conveyed object names and types, although sensory and size information were often missing. Route knowledge was not applicable in most cases, but the one text map with defined routes (Turn-by-Turn Directions) omitted all features outside the route, as well as the alternative routes presented on the visual map. Survey knowledge could not be evaluated for any of the failing maps due to missing features.

## Discussion

The results show that many legacy text map methods (e.g., tables and turn-by-turn directions) fail the MEP Framework evaluation and are not WCAG compliant. The biggest failures were the text map missing features, and failing to communicate the spatial relationships between features. Several other passing text alternatives are provided, although it is critical that a holistic, and systematic, map creation approach be performed to ensure equivalent purpose to a visual map. The “primary purpose” or “point” of a map is important as a piece of the textual information, but is insufficient on its own. There could be use cases for a map that are outside the imagination of the describer but are possible using the visual map, and these cases need to be accounted for in the description as well to establish equivalence of information. For example, originally, the graphic for the audio description and MUD Map contained the names of small cities, but these were not provided in the text description. In this case, the visual map was updated to remove the city names to match the text description.

Aside from everyday uses, having equally accessible maps is critical for BLVIs to participate in map-dependent professions. Although the number of BLVIs in map-dependent professions is generally unknown, fewer than a fifth of BLVIs even have used a tactile map (Bleau et al.), and those who have used a map report using fewer than one per year, on average (Biggs, Pitcher-Cooper, et al.). This is in contrast to sighted users, who view more than 300 maps a year on average (Savino et al.). Map research exclusively depending on BLVI feedback is similar to asking monastic scribes, who only have access to quill and parchment, what features they want in their word processing software. The MEP framework can supplement BLVI feedback to ensure all information is present in the map. Besides common uses, numerous lucrative professions also require the use of maps (Department of Geography and Geosciences, Salisbury University), and the widespread lack of comprehensive text maps is keeping people from these jobs just because they are BLVIs. Map tool creators, such as those evaluated in the paper by Biggs and colleagues Biggs, Coughlan, et al., need to use one of the passing methods to be considered compliant with the new global accessibility legislation and create equality in the cartographic-dependent professions. Government agencies need to evaluate map solutions using the MEP Framework, along with the other WCAG criteria outlined in Biggs, Coughlan, et al., to ensure BLVIs and other disabled people are able to participate in civic projects, access emergency information, understand utility maps, and work in professions depending on these maps. State and local government documentation needs to be updated based on the results of this evaluation (Colorado OIT-GIS; Minnesota IT Services). Both the European Accessibility Act and The Public Sector Bodies Accessibility Regulation should also remove the explicit exception for thematic maps, since all three passing maps in this evaluation were thematic (European Commission; *The Public Sector Bodies (Websites and Mobile Applications) (No. 2) Accessibility Regulations 2018*).

## Conclusions

This paper developed the Map Equivalent-Purpose (MEP) Framework to systematically assess whether text-based maps achieve the equivalent purpose of visual maps, as required by WCAG 1.1.1. Eight text-based formats were evaluated, with only interactive Audiom maps, MUD maps, and audio descriptions meeting the framework's criteria; tables, turn-by-turn directions, nearby-address searches, Google Maps interactive alternative text, and short alt text failed, indicating these legacy formats are insufficient for WCAG compliance. While expert analysis demonstrates the framework's utility, further empirical validation is needed through controlled spatial-knowledge tasks to test the emerging Equivalent Purpose Theory, which posits that a text map achieves equivalent purpose when it supports the same generalized purpose, spatial information, and spatial-relationship understanding as its visual counterpart. Limitations include reliance on visual baselines, and the absence of temporal and chat-based interpretive map representations (Froehlich et al.; Jain et al.). Future research should explore non-visual sensory maps, extend the framework to map creation tools (Clepper et al.), and adapt the evaluation approach to other digital spatial representations (e.g., bar charts, 3D objects, Sankey diagrams).

Overall, the MEP Framework provides a replicable template for assessing geographic text maps, clarifies the concept of equivalent purpose, expands W3C technique G92, highlights the insufficiency of legacy methods, and offers guidance for developing accessible map representations that support BLVIs' participation in cartographic-dependent professions.

*Table 6: metadata*

| Section | Content |
|---|---|
| Supplemental Data | This is a link to a Google Sheet with the MEP Framework template and coded data.<br>A permanent archive of the above MEP Framework template can be found at: (Biggs)<br><br>Here is a Github repository where you can comment on modifications to the MEP Framework. |
| Disclosure of Interest | Mr. Biggs is the founder and CEO of XR Navigation. Dr. Brett Oppegaard is the founder of AccessHound. Dr. Nicholas A. Giudice is a founder of UNAR Labs. All three companies work to make digital maps accessible. |
| Funding | TThis work was funded under National Institute on Disability, Independent Living, and Rehabilitation Research Grant No. 90REGE0018, and NIH Grant No. 1 R44 EY036316-01A1. |
| Acknowledgements | We would like to thank both Kira Spivey and Isaiah Kenny for evaluating the eight text maps. |

## Works Cited

Accessibility Guidelines Working Group. *Technique G92:providing Long Description for Non-Text Content That Serves the Same Purpose and Presents the Same Information*. World Wide Web Consortium, https://www.w3.org/WAI/WCAG21/Techniques/general/G92.

*AirNow*. Office of Air Quality Planning and Standards, 2021, https://www.airnow.gov/.

Arundel, Steven T., and W. Li. "The Evolution of Geospatial Reasoning, Analytics, and Modeling." *The Geographic Information Science & Technology Body of Knowledge*, edited by John P. Wilson, 2021, https://doi.org/10.22224/gistbok/2021.3.4.

Aziz, Nida, et al. "Planning Your Journey in Audio: Design and Evaluation of Auditory Route Overviews." *ACM Transactions on Accessible Computing*, vol. 15, no. 4, ACM New York, NY, 2022, pp. 1–48, https://doi.org/10.1145/3531529.

Bank of America Corporation. *Bank of America Financial Centers and ATMs*. 2025, https://locators.bankofamerica.com/?q=94115.

Biggs, Brandon, Christopher Toth, et al. "Evaluation of a Non-Visual Auditory Choropleth and Travel Map Viewer." *International Conference on Auditory Display*, International Conference on Auditory Display, 2022, https://doi.org/10.21785/icad2022.027.

Biggs, Brandon, Charity Pitcher-Cooper, et al. "Getting in Touch with Tactile Map Automated Production: Evaluating Impact and Areas for Improvement." *Journal on Technology and Persons with Disabilities*, vol. 10, 2022, https://pmc.ncbi.nlm.nih.gov/articles/PMC10065749/.

Biggs, Brandon, James M. Coughlan, et al. “Systematically Evaluating Digital Map Tools Based on the WCAG.” *Journal on Technology and Persons with Disabilities*, vol. 13, 2025, http://hdl.handle.net/20.500.12680/qn59qf178.

Biggs, Brandon, Lena Yusim, et al. *The Audio Game Laboratory: Building Maps from Games*. 2018, https://icad2018.icad.org/wp-content/uploads/2018/06/ICAD2018_paper_51.pdf.

Bleau, Maxime, et al. “International Prevalence of Tactile Map Usage and Its Impact on Navigational Independence and Well-Being of People with Visual Impairments.” *Scientific Reports*, vol. 15, no. 1, Nature Publishing Group, 2025, p. 27245, https://doi.org/10.1038/s41598-025-08117-9.

Cassidy, Johny. “Complex Images: How to Write Text Descriptions for Graphs, Charts, Maps or Infographics That Convey Detailed Factual Information.” *How to Write Text Descriptions (Alt Text)*, BBC, 2024, https://www.bbc.co.uk/gel/features/complex-images.

Clementini, Eliseo, et al. “Modelling Topological Spatial Relations: Strategies for Query Processing.” *Computers & Graphics*, vol. 18, no. 6, Elsevier, 1994, pp. 815–22, https://doi.org/10.1016/0097-8493(94)90007-8.

Clepper, Gina, et al. “" What Would i Want to Make? Probably Everything": Practices and Speculations of Blind and Low Vision Tactile Graphics Creators.” *Proceedings of the 2025 CHI Conference on Human Factors in Computing Systems*, 2025, pp. 1–16, https://ej-mcdonnell.github.io/tactileGraphics.pdf.

*ClickAndGo Wayfinding*. ClickAndGo Wayfinding Maps, 2014, https://www.clickandgomaps.com/.

Colorado OIT-GIS. *Colorado GIS Accessibility Guidelines*. 2024, https://gis.colorado.gov/accessibility/?page=Home.

Conway, Megan, et al. "Audio Description: Making Useful Maps for Blind and Visually Impaired People." *Technical Communication*, vol. 67, no. 2, Society for Technical Communication, 2020, pp. 68–86, https://www.stc.org/techcomm/2020/04/28/audio-description-making-useful-maps-for-blind-and-visually-impaired-people/.

Coughlan, James M., et al. "Non-Visual Access to an Interactive 3D Map." *Computers Helping People with Special Needs: 18th International Conference, ICCHP-AAATE 2022, Lecco, Italy, July 11–15, 2022, Proceedings, Part i*, Springer, 2022, pp. 253–60, https://doi.org/10.1007/978-3-031-08648-9_29.

Department of Geography and Geosciences, Salisbury University. *Careers in Geography and Geosciences: A World of Possibilities*. https://www.salisbury.edu/academic-offices/science-and-technology/geography-and-geosciences/_files/careers-in-geosciences.pdf.

Desrosiers, Caroline. *Alt Text Character Limit*. 2021, https://github.com/w3c/wcag/discussions/4047.

European Commission. *European Accessibility Act*. 2017, https://eur-lex.europa.eu/legal-content/EN/TXT/?uri = CELEX%3A32019L0882.

Foster, M. "Statistical Mapping (Enumeration, Normalization, Classification)." *The Geographic Information Science & Technology Body of Knowledge (2nd Quarter 2019 Edition)*, edited by John P. Wilson, University Consortium for Geographic Information Science, 2019, https://doi.org/10.22224/gistbok/2019.2.2.

Froehlich, Jon E., et al. "StreetViewAI: Making Street View Accessible Using Context-Aware Multimodal AI." *Proceedings of the 2025 ACM Symposium on User Interface Software and Technology*, 2025, https://doi.org/10.1145/3746059.3747756.

Giudice, Nicholas A., et al. "Cognitive Mapping Without Vision: Comparing Wayfinding Performance After Learning from Digital Touchscreen-Based Multimodal Maps Vs. Embossed Tactile Overlays." *Frontiers in Human Neuroscience*, vol. 14, Frontiers Media SA, 2020, https://doi.org/10.3389/fnhum.2020.00087.

Google. *Accessibility in Google Maps*. 2019, https://support.google.com/maps/answer/6396990?co = GENIE.Platform%3DDesktop&hl = en.

Griatch. *Evennia Python MUD/MU Creation System*. 2025, https://github.com/evennia/evennia.

Guidero, E. "Typography." *The Geographic Information Science & Technology Body of Knowledge (3rd Quarter 2017 Edition)*, edited by John P. Wilson, University Consortium for Geographic Information Science, 2017, https://doi.org/10.22224/gistbok/2017.3.2.

Harvey, Francis. "That Is Not a Pipe, Mon. Magritte. But, Is This Not a Map? Some Questions about Map Correspondence Inspired Through Barbara Buttenfield." *Cartography and Geographic Information Science*, vol. 51, no. 5, Taylor & Francis, 2024, pp. 649–58, https://www.tandfonline.com/doi/abs/10.1080/15230406.2024.2311265.

Hennig, Sabine, et al. "Accessible Web Maps for Visually Impaired Users: Recommendations and Example Solutions." *Cartographic Perspectives*, no. 88, 2017, pp. 6–27, https://doi.org/10.14714/cp88.1391.

Jain, Gaurav, et al. *SceneScout: Towards AI Agent-Driven Access to Street View Imagery for Blind Users*. 2025, https://arxiv.org/abs/2504.09227.

Juan-Armero, Sergio, and Sergio Luján-Mora. "Using SVG to Develop Web Maps for People with Visual Disabilities." *Enfoque UTE*, vol. 10, no. 2, Universidad Tecnológica Equinoccial, 2019, pp. 90–106, https://doi.org/10.1002/9780470379424.ch8.

KMSOC. *Comment 3*. 2024, https://github.com/atbcb/ICTTestingBaseline/issues/477#issuecomment-2223501226.

Lapaine, Miljenko, et al. "Definition of the Map." *Advances in Cartography and GIScience of the International Cartographic Association*, vol. 3, 2021, p. 9, https://ica-adv.copernicus.org/articles/3/9/2021/ica-adv-3-9-2021.pdf.

Li, Zhilin, and Peizhi Huang. "Quantitative Measures for Spatial Information of Maps." *International Journal of Geographical Information Science*, vol. 16, no. 7, Taylor & Francis, 2002, pp. 699–709, https://www.tandfonline.com/doi/abs/10.1080/13658810210149416.

Loeliger, Esther, and Tony Stockman. "Wayfinding Without Visual Cues: Evaluation of an Interactive Audio Map System." *Interacting with Computers*, vol. 26, no. 5, OUP, 2014, pp. 403–16, https://doi.org/10.1093/iwc/iwt042.

Logan, Thomas. *Accessible Maps on the Web*. Equal Entry, 2018, https://equalentry.com/accessible-maps-on-the-web/.

---. *Let's Make Maps Widely Accessible*. 2018, https://www.youtube.com/watch?v = 0gB83fkfD8Y&list = PLn7dsvRdQEfEnBxpVztmJ8KCKNJ_P-hR6.

*Map Definition*. Merriam Webster, https://www.merriam-webster.com/dictionary/map.

---. Wikipedia, https://en.wikipedia.org/wiki/Map.

*mapbox-gl-accessibility*. Mapbox, https://github.com/mapbox/mapbox-gl-accessibility.

McCall, Karen, and Beverly Chagnon. “Rethinking Alt Text to Improve Its Effectiveness.” *18th International Conference, ICCHP-AAATE 2022*, ICCHP-AAATE; Springer, 2022, https://link.springer.com/chapter/10.1007/978-3-031-08645-8_4.

Minnesota IT Services. *Map Accessibility*. Minnesota IT Services, https://mn.gov/mnit/about-mnit/accessibility/maps/.

*PO Locator*. USPS, https://tools.usps.com/find-location.htm.

Roth, Robert E. “Visual Variables.” *International Encyclopedia of Geography: People, the Earth, Environment and Technology*, John Wiley & Sons, Ltd, 2017, pp. 1–11, https://learn.uni-med.net/pluginfile.php/17434/mod_resource/content/2/Roth_2016_VisualVariables.pdf.

Savino, Gian-Luca, et al. “MapRecorder: Analysing Real-World Usage of Mobile Map Applications.” *Behaviour & Information Technology*, vol. 40, no. 7, Taylor & Francis, 2021, pp. 646–62, https://doi.org/10.1080/0144929x.2020.1714733.

Shull, Chris J. *Improved Accessibility in the Maps JavaScript API*. Google, 2021, https://cloud.google.com/blog/products/maps-platform/improved-accessibility-maps-javascript-api.

Siegel, Alexander W., and Sheldon H. White. “The Development of Spatial Representations of Large-Scale Environments.” *Advances in Child Development and Behavior*, vol. 10, Elsevier, 1975, pp. 9–55, https://doi.org/10.1016/s0065-2407(08)60007-5.

Sloan, David. “Accessible Digital Map Experiences: A Mountain Climb or a Walk in the Park?” *TPGi Blog*, 2020, https://www.tpgi.com/accessible-digital-map-experiences/.

Song, Y. *Time*. Edited by John P. Wilson, 2019, pp. The Geographic Information Science & Technology Body of Knowledge (4th Quarter 2019 Edition), https://doi.org/10.22224/gistbok/2019.4.7.

The Mud Connector. *Getting Started: Welcome to Your First MUD Adventure*. 2013, http://www.mudconnect.com/mud_intro.html.

*The Public Sector Bodies (Websites and Mobile Applications) (No. 2) Accessibility Regulations 2018*. United Kingdom Crown, 2018, https://www.legislation.gov.uk/uksi/2018/952/made.

UniDescription. *The UniDescription Project*. http://www.unidescription.org/.

*UniDescription Academy*. UniDescription, https://unidescription.org/unid-academy.

*United States COVID-19 Cases, Deaths, and Laboratory Testing (NAATs) by State, Territory, and Jurisdiction*. Centers for Disease Control and Prevention, https://covid.cdc.gov/covid-data-tracker/#cases_newcaserateper100k.

U.S. Department of Justice Civil Rights Division. *Fact Sheet: New Rule on the Accessibility of Web Content and Mobile Apps Provided by State and Local Governments*. 2024, https://www.ada.gov/resources/2024-03-08-web-rule/.

W3C Web Accessibility Initiative. *Complex Images*. https://www.w3.org/WAI/tutorials/images/complex/, Jan. 2022, https://www.w3.org/WAI/tutorials/images/complex/.

W3C Web Accessibility Initiative. *Web Accessibility Laws & Policies*. 2023, https://www.w3.org/WAI/policies/.

*Web Content Accessibility Guidelines (WCAG) Overview*. W3C, 2018, https://www.w3.org/WAI/standards-guidelines/wcag/.

*Welcome to XR Navigation!* XR Navigation, https://xrnavigation.io/.

White, T. "Symbolization and the Visual Variables." *The Geographic Information Science & Technology Body of Knowledge (2nd Quarter 2017 Edition)*, edited by John P. Wilson, University Consortium for Geographic Information Science, 2017, https://doi.org/10.22224/gistbok/2017.2.3.

Williams, I. *How Blind Players Made a Text-Only RPG More Accessible*. 2016, https://waypoint.vice.com/en_us/article/kwz77n/how-blind-players-made-a-text-only-rpg-more-accessible.

World Wide Web Consortium. *Understanding SC 1.1.1:non-Text Content (Level a)*. https://www.w3.org/WAI/WCAG21/Understanding/non-text-content.html.

Zhao, H., et al. "Data Sonification for Users with Visual Impairment: A Case Study with Georeferenced Data." *ACM Transactions on Computer-Human Interaction (TOCHI)*, vol. 15, no. 1, 2008, pp. 1–28, https://doi.org/10.1145/1352782.1352786.